\documentstyle[preprint,aps]{revtex}
\begin{document}
\draft
\title{Cylindrical solutions in braneworld gravity}
\author{S.~Khoeini-Moghaddam$^{a}\thanks{Electronic
address:~saloumeh@mehr.sharif.edu}$ and
M.~Nouri-Zonoz$^{b,c}\thanks{Electronic
address:~nouri@khayam.ut.ac.ir, Corresponding author}$}
\address{$^{a}$Department of Physics, Sharif University of Technology,
P O Box 11365-9161 Tehran, Iran \\
$^{b} $ Department of Physics, University of Tehran, End of North Karegar St.,
14395-547 Tehran, Iran
\\$^{c}$ Institute for Studies in Theoretical Physics and
Mathematics, P O Box 19395-5531 Tehran, Iran} 
\maketitle
\begin{abstract}
In this article we investigate exact cylindrically symmetric
solutions to the modified Einstein field equations in the brane world
gravity scenarios. It is shown that for the special choice of the equation of
state $2U+P=0$ for the dark energy and dark pressure, the solutions found
could be considered formally as solutions of the Einstein-Maxwell equations in 4-D
general relativity.
\end{abstract}
\pacs{PACS No.04.50+h , 04.40.Nr}
%%%%%%%%%%%%%%%%%%%%%%%%%%%%%%%%%%%%%
%----------------------begin section-------------------------
\section{Introduction}
Since Einstein's formulation of general theory of relativity ,
extra dimensions have always played a profound role in studies
trying to unify gravity with other forces of nature. The very well
known example is the Kaluza-Klein 5-dimensional theory whose main
goal was the unification of gravitational and electromagnetic
interactions via compactified fifth dimension. In recent years
string inspired non-compactified extra dimensions have raised a
lot of enthusiasm in the so called {\it brane world gravity}
scenarios. In these scenarios, extra dimensions being large need
not be compactified, instead we expect imprints of embedding and
free gravitational field (in the 5-dimensional bulk spacetime of
scenarios with one extra dimension) to show up in modified
dynamical equations on the brane which are different from the
usual four dimensional Einstein equations. Consequently, when we
confine ourselves to the 3-brane, exact solutions to these
modified field equations are expected to be either different from
those in the usual four-dimensional general relativity or, if they
are formally the same, to be reinterpreted in terms of some of the
characteristics of the extra dimension. In 5-dimensional case,
exact radial solutions with one and two parameters, are studied in
\cite{Harma} where it is shown that there are different solutions
according to the different equations of state for dark radiation
and dark pressure. Gravitational collapse of matter, producing
black holes on a brane, is studied in \cite{Chamb}. There it is
shown that a five dimensional {\it black string} solution
intersects the brane in a Schwarzschild black hole. In the same
scenario it is also shown that the two parameter
Reissner-Nordstrom type solution could be interpreted in terms of
mass and another parameter attributed to a characteristic of the
bulk space \cite{Dad}. In this paper, using the modified Einstein
equations on the brane, we study exact solutions on the brane with
cylindrical symmetry. Actually what we will show is another
manifestation of the fact that any solution of the
Einstein-Maxwell equations (with traceless energy-momentum tensor)
in 4-dimensional general relativity could be interpreted as a {\it
vacuum} solution of brane-world 5-dimensional gravity \cite{Dad}.
%----------------------begin section-------------------------
\section{Einstein field equations on the brane}
A covariant generalization of Randall-Sundrum model\cite{RS1,RS2}
is given in \cite{Shiro} (the so called SMS braneworld \footnote{Note that there are reformulations 
of the SMS braneworld in the literature with respect to different aspects of that formulation \cite{tav,ali}.} ) where the 4-dimensional world is described
by a 3-brane $(M,q_{\mu\nu})$ as a fixed point of $Z_2$ symmetry in a 
5-dimensional space-time
$({\cal{M}},g_{\mu\nu})$. Choosing the 5-th coordinate $\chi$ such that the brane
coincides with the hypersurface $\chi=0$, the 5-dimensional metric could be written as;
\begin{equation}\label{1.0}
ds^2=(n_{\mu}n_{\nu}+q_{\mu\nu})dx^{\mu}dx^{\nu}
\end{equation}
where $n_\mu$ is the unit vector normal to the brane i.e $n_\mu dx^\mu=d\chi$ and
$q_{\mu\nu}$ is the induced metric on the brane.
Assuming that the matter is confined to the brane, the 5-dimensional 
Einstein equations read;
\begin{eqnarray}\label{1.1}
{ }^{(5)}{G}_{\mu\nu}&=&\kappa_{(5)}^{2}{ }^{(5)}{T}_{\mu\nu}\nonumber\\
{}^{(5)}{T}_{\mu\nu}&=&-\Lambda_{(5)}g_{\mu\nu}+\delta(\chi)[-\lambda_{b}q_{\mu\nu}
+{}^{(4)}T_{\mu\nu}]
\end{eqnarray}
where $\Lambda_{(5)}$ is the bulk cosmological constant and
$\lambda_{b}$ is the vacuum energy on the brane or the brane
tension in the bulk space. Since we assumed that the matter is
confined to the brane, the 4-dimensional matter field Lagrangian
determines the ${ }^{(4)}T_{\mu\nu}$ such that ${}^{(4)}T_{\mu\nu}n^\mu=0$.
Using the Gauss-Codacci equations, along with the $Z_2$ symmetry, one could
 project 5-dimensional curvature equations along the
4-dimensional hypersurface coincident with the brane. The effective
Einstein field equations on the brane are found to be \cite{Shiro};
\begin{equation}\label{1.2}
{ }^{(4)}G_{\mu\nu}=-\Lambda q_{\mu\nu}+\kappa_{(4)}^{2}{
}^{(4)}T_{\mu\nu}+\kappa_{(5)}^{4}S_{\mu\nu}-\mathcal{E}_{\mu\nu}
\end{equation}
where
\begin{eqnarray}
S_{\mu\nu}&=&\frac{1}{12}{ }^{(4)}T{
}^{(4)}T_{\mu\nu}-\frac{1}{4}{ }^{(4)}T_{\mu}^{\alpha}{
}^{(4)}T_{\alpha\nu} +\frac{1}{24}q_{\mu\nu}(3{
}^{(4)}T^{\alpha\beta}{ }^{(4)}T_{\alpha\beta}-{
}^{(4)}T^2)\label{1.3}
\end{eqnarray}

\begin{eqnarray}\label{1.5}
\kappa_{(4)}^{2}&=&8\pi G_N\nonumber\\
G_N&=&\frac{\kappa_{(5)}^{4}\lambda_{b}}{48 \pi}\\
\Lambda&=&\frac{1}{2}\kappa_{(5)}^{2}(\Lambda_{(5)}+
\frac{1}{6}\kappa_{(5)}^{2}\lambda_{b}^2)\nonumber
\end{eqnarray}
and
\begin{equation}\label{projwyel}
{\mathcal E}_{\mu\nu}={
}^{(5)}{C}^{\alpha}_{\beta\gamma\delta}n_{\alpha}n^{\gamma}q_{\mu}^{\beta}q_{\nu}^{\delta}
\end{equation}
is the transmitted projection of the bulk Weyl tensor\footnote{ To remind 
the reader, the n-dimensional ($n\geq 3$) Weyl tensor is defined by,
\begin{eqnarray}
{}^{(n)}{C}_{\alpha\beta\gamma\delta}= {}^{(n)}{R}_{\alpha\beta\gamma\delta} 
+{1\over n-2}(g_{\alpha\delta}{}^{(n)}{R}_{\beta\gamma} + g_{\beta\gamma}{}^{(n)}{R}_{\delta\alpha}
 - g_{\alpha\gamma}{}^{(n)}{R}_{\delta\beta} - g_{\beta\delta}{}^{(n)}{R}_{\gamma\alpha})
\nonumber  \\+ {1\over (n-1)(n-2)}(g_{\alpha\gamma}g_{\beta\delta} - g_{\alpha\delta}g_{\gamma\beta}){}^{(n)}{R}. \nonumber
\end{eqnarray} }.
Note that it is symmetric and traceless. Equation (\ref{1.2}) is the
modified Einstein field equation on the brane due to the bulk
effects. Compared with the usual 4-dimensional Einstein field
equations, there are two main modifications: the first one is the
presence of $S_{\mu\nu}$ term which is quadratic in ${
}^{(4)}T_{\mu\nu}$. Comparison of the second and third terms in
the right hand side of (\ref{1.2}) shows that it is important in
high energy limit and dominates when the energy density $\rho \gg
\lambda_b$ and is negligible for $\rho \ll \lambda_b$. The second
correction is due to ${\mathcal{E}}_{\mu\nu}$, the projection of
the bulk Weyl tensor on the brane. From the viewpoint of a
brane-observer the former is local and the latter is non-local.
One recovers the usual Einstein field equations, by taking the
limit $\kappa_{(5)}\rightarrow 0$ while keeping $G$ finite. Though
there is a  point to be made here, according to (\ref{1.5}) it is
impossible to define Newton's gravitational constant $G_N$ without
an unambiguous definition of $\lambda_b$\cite{csaki}. Using
equation (\ref{1.2}), the 4D contracted Bianchi identity
${\mathcal{D}}^{\nu}{ }^{(4)}G_{\mu\nu}=0$ along with the
conservation of ${ }^{(4)}T_{\mu\nu}$,
 ${\mathcal{D}}^{\nu}{ }^{(4)}T_{\mu\nu}=0$ lead to the following constraint;
%on the covariant derivatives of ${\mathcal{E}}_{\mu\nu}$ and $S_{\mu\nu}$;
\begin{equation}\label{1.7}
  {\mathcal{D}}^{\mu}{\mathcal{E}}_{\mu\nu}=\kappa_{(5)}^{4}{\mathcal{D}}^{\mu}S_{\mu\nu},
\end{equation}
Where ${\mathcal{D}}^{\mu}$ is the covariant derivative with respect to $q_{\mu\nu}$.
Being symmetric and traceless, ${\mathcal{E}}_{\mu\nu}$ could be decomposed
irreducibly with respect to a chosen 4- velocity vector field
$u^{\mu}$ as \cite{Maar1};
\begin{equation}\label{1.8}
{\mathcal{E}}_{\mu\nu}=-\kappa^{4}[\textit{U}(u_{\mu}u_{\nu}+\frac{1}{3}h_{\mu\nu})+
\textit{P}_{\mu\nu}+2\textit{Q}_{(\mu }u_{\nu )}]
\end{equation}
where $h_{\mu\nu}=q_{\mu\nu}+u_{\mu}u_{\nu}$ is the projection
tensor orthogonal to $u^{\mu}$, $\kappa=\kappa_{(5)}/\kappa_{(4)}$
and
\begin{eqnarray}\label{1.9}
\textit{U}&=&-\kappa^{-4}{\mathcal{E}}_{\mu\nu}u^{\mu}u^{\nu}\nonumber\\
\textit{Q}_{\mu}&=&\kappa^{-4}h_{\mu}^{\alpha}{\mathcal{E}}_{\alpha\beta}u^{\beta} \\
\textit{P}_{\mu \nu }&=&-k^{-4} [h_{(\mu } ^{\alpha }h_{\nu
)}^{\beta }-\frac{1}{3}h_{\mu \nu }h^{\alpha \beta }]
{\mathcal{E}}_{\alpha \beta }\nonumber
\end{eqnarray}
$\textit{U}$ is an effective nonlocal energy density on the
brane, arising from the bulk free gravitational field. Note that
this nonlocal energy density need not be positive (in fact
$\textit{U}$ contributes to tidal acceleration on the brane in the
off-brane direction \cite{Maar1}) and actually being negative is 
more consistent with the Newotonian picture in which gravitational field 
carries negative energy . Also $\textit{P}_{\mu \nu }$ and $\textit{Q}_{\mu}$
are respectively the effective nonlocal anisotropic stress and energy 
flux on the brane, both arising from the free gravitational field in the bulk.
%------------------------------------begin section ------------------------------
\section{Static Cylindrically Symmetric Solution}
In this section we solve the 4-dimensional modified Einstein
equation (\ref{1.2}) in vacuum ($T_{\mu\nu}=0$) for a
static, cylindrically symmetric case. In the vacuum
$T_{\mu\nu}$ and consequently $S_{\mu\nu}$ vanish and the constraint (\ref{1.7})
becomes
\begin{equation}\label{1.19}
{\mathcal{D}}^\mu {\mathcal{E}}_{\mu\nu}=0
\end{equation}
One can Choose $\Lambda =0$ by taking $\Lambda_{(5)} =
-1/6\kappa^2_{(5)}\lambda_b^2$ so that the effective Einstein
equations on the brane reduce to
\begin{equation}\label{1.30}
{ }^{(4)}R_{\mu\nu} = -\mathcal{E}_{\mu\nu}
\end{equation}
As pointed out in \cite{Dad} the set of equations
(\ref{1.19}-\ref{1.30}) form a closed system of equations for
static solutions on the brane. In particular Einstein-Maxwell
solutions (with traceless energy-momentum
tensor) in 4-dimensional general relativity could be properly interpreted as 
{\it vacuum} brane world solutions.\\
Representing nonlocal effects, one could  specify ${\mathcal{E}}_{\mu\nu}$ 
in an inertial frame at any point on the brane where $u^\mu = \delta^\mu_0$. 
In static spacetimes this corresponds to comoving frames along the timelike 
Killing vector field for which \cite{Dad}
\begin{equation}\label{1.31}
\textit{Q}_\mu=0\;\;\;\;\; , \;\;\;\;\;h_{\mu}^{\nu} = {\rm diag} (0,1,1,1)
\end{equation}
and the constraint for ${\mathcal{E}}_{\mu\nu}$ takes the form
\begin{equation}\label{1.10}
\frac{1}{3}\tilde{\nabla}_\mu\textit{U}+\frac{4}{3}\textit{U}A_\mu+
\tilde{\nabla}^\nu\textit{P}_{\mu\nu}+A^\nu\textit{P}_{\mu\nu}=0
\end{equation}
where $\tilde{\nabla}_\mu=h_{\mu}^{\nu}{\mathcal{D}}_\nu$ is the projected covariant 
derivative orthogonal to $u^\mu$  and
$A_\mu=u^\nu{\mathcal{D}}_\nu u_\mu$ is the 4-acceleration. In the
static cylindrically symmetric case we may choose
\begin{equation}\label{A}
A_\mu =A(\rho)\rho_\mu
\end{equation}
and
\begin{equation}\label{P}
  \textit{P}_{\mu\nu}=P(\rho)(\rho_\mu\rho_\nu-\frac{1}{3}h_{\mu\nu}).
\end{equation}
where $A(\rho)$  and $P(\rho)$ are scalar functions and $\rho_\mu$ is the radial unit 
vector. Now in the (inertial) comoving frame the projected bulk Weyl tensor takes the 
form;
\begin{equation}\label{1.22}
{\mathcal{E}}^{\mu}_{\nu}= -\kappa^4 {\rm diag}(-U , {1\over 3}(U+2P) , {1\over 3}(U-P) , {1\over 3}(U-P))
\end{equation}
Choosing the following general form for a static cylindrically symmetric
line element on the brane,
\begin{equation}\label{1.11}
ds^2=-e^{2f(\rho)}dt^2+e^{-2f(\rho)}e^{2K(\rho)}(d\rho^2+dz^2)+e^{-2f(\rho)}W(\rho)^2d\phi^2
\end{equation}
substituting it into the modified gravitational field equations
 (\ref{1.30}) and using (\ref{1.22}) we end up with,
\begin{eqnarray}
-\frac{1}{W}(-2f''W-2f'W'+f'^2W+K''W+W'')&=&-\kappa^4\textit{U}
e^{-2f}e^{2K}\label{1.12}\\
-\frac{1}{W}(f'^2W-K'W')&=&\frac{-\kappa^4}{3}(\textit{U}+2\textit{P})e^{-2f}e^{2K}\label{1.13}\\
\frac{1}{W}(f'^2W-K'W'+W'')&=&\frac{-\kappa^4}{3}(\textit{U}-\textit{P})e^{-2f}e^{2K}\label{1.14}\\
(f'^2+K'')&=&\frac{-\kappa^4}{3}(\textit{U}-\textit{P})e^{-2f}e^{2K}\label{1.15}
\end{eqnarray}
The equation (\ref{1.19})  could be written explicitly  as
\begin{equation}\label{1.36}
\textit{U}^{'}-(4f^{'}-2K^{'}-\frac{2W^{'}}{W})\textit{U}=0
\end{equation}
From (\ref{1.13}) and (\ref{1.14}), we obtain
\begin{equation}\label{1.0015}
  \frac{W''}{W}=\frac{-\kappa^4}{3}(2\textit{U}+\textit{P})e^{-2f}e^{2K}.
\end{equation}
To solve the above equations we choose the following equation of state
\begin{equation}\label{1.015}
    2\textit{U}+\textit{P}=0
\end{equation}
There is no physical restriction that implies this choice of equation of state, but it is not
an easy task to solve the above equations in their general form, i.e for arbitrary $U(\rho)$ 
and $P(\rho)$, even in the spherical case \cite{Dad}. On the other hand it is 
interesting to note that the above equation of state in
terms of the components of the projected Weyl tensor reads;
\begin{equation}\label{1.35}
{\mathcal{E}}^{0}_{0}+{\mathcal{E}}^{3}_{3}= 0
\end{equation}
This establishes, once more and now for cylindrical symmetry, the connection stated 
above between solutions of the Einstein-Maxwell equations in general relativity 
and those of the vacuum brane world. For, (\ref{1.35}) is the equation satisfied by 
the energy momentum tensor of Maxwell fields leading to  known cylindrically 
symmetric solutions to the Einstein-Maxwell equations \cite{Steph}. We will discuss 
this in more detail in the next section.\\
By the above choice of the equation of state and from (\ref{1.0015}) we find
$W''=0$ or $W=a\rho+b$, which by appropriate scaling of
coordinates  means either $W=\rho$ or $W=const$. Choosing $W=\rho$,
 (\ref{1.14}) and (\ref{1.15}) lead to the following solutions for $e^{2K}$,
\begin{equation}\label{1.16}
e^{2K}=\rho^{2m^2}
\end{equation}
or
\begin{equation}\label{1.161}
e^{2K}=1
\end{equation}
Using the first solution and solving equations (\ref{1.12}-\ref{1.36}) we end up
with
\begin{eqnarray}
e^{2f}=\frac{4m^2}{(c_1\rho^{-m}+c_2\rho^{m})^2}\\
U(\rho)=\frac{16m^2c_1c_2}{\kappa^4}\frac{\rho^{-2-2m^2}}{(c_1\rho^{-m}+c_2\rho^{m})^4}
\end{eqnarray}
where $m$, $c_1$ and $c_2$ are constants and $c_1c_2 < 0 $. This last condition on the constants $c_1$ and $c_2$ could be infered from the comparison of this solution with the corresponding Einstein-Maxwell solution (refer to the next section).
For the latter choice of $e^{2K}$ we arrive at
\begin{eqnarray}
e^{2f}=\frac{1}{(c_1\ln\rho+c_2)^2}\\
U(\rho)=-\frac{c_1^2}{\kappa^4\rho^2(c_1\ln\rho+c_2)^4}
\end{eqnarray}
Now setting $W=1$ and solving Einstein field equations along with the constraint 
equation (\ref{1.36}) we find
\begin{eqnarray}\label{2.1}
K&=&a\rho+b\nonumber\\
f&=&-\ln(\rho)\\
\textit{U}&=&-{\kappa^4}\frac{e^{-2(a\rho+b)}}{\rho^4}\nonumber
\end{eqnarray}
%--------------------------------begin section---------------------------------
\section{Relation to Einstein-Maxwell solutions }
As pointed out previously, the brane world solutions found in the 
last section could be treated as the solutions to the 
Einstein-Maxwell equations in general relativity. So one could use
this analogy to find expressions for the corresponding electromagnetic 
fields in terms of the dark energy $U$ (or dark pressure $P$).
We note that the solutions (26-32) are static cylindrically symmetric 
electrovacs discussed in the exact solutions literature (see
section 22.2 of  \cite{Steph} and references therin).
Assuming that the electromagnetic fields inherit the metric symmetry and
are orthogonal to the orbits of the two
dimentional orthogonally transitive group, the
vector potential could be written as \cite{Steph}
\begin{equation}
A_{\alpha}dx^{\alpha}={\mathcal{P}}(\rho)dt+{\mathcal{Q}}(\rho)d\phi.
\end{equation}
This vector potential corresponds to a magnetic field along the
z-direction and an electric field along the $\rho$-direction. The 
non-zero components of $F_{\mu\nu}$ are $F_{01}={\mathcal{P}}'$ and
$F_{13}=-{\mathcal{Q}}'$ so that the corresponding energy-momentum tensor 
components $T_{\mu}^{\nu}$ are given by;
%\footnote{$T_{\mu\nu}=-F_{\mu\gamma}F_{\nu}^{\gamma}+
%\frac{1}{4}g_{\mu\nu}F_{\gamma\delta}F^{\gamma\delta}$}
%from this vector potential we get
\begin{eqnarray}
T_{0}^{0}&=&-\frac{1}{2}e^{-2K}{\mathcal{P}}^{'2}-
\frac{1}{2}e^{4f}e^{-2K}W^{-2}{\mathcal{Q}}^{'2}\label{5.2}\\
T_{1}^{1}&=&-\frac{1}{2}e^{-2K}{\mathcal{P}}^{'2}+
\frac{1}{2} e^{4f}e^{-2K}W^{-2}{\mathcal{Q}}^{'2}\label{5.3}\\
T_{2}^{2}&=&\frac{1}{2}e^{-2K}{\mathcal{P}}^{'2}-
\frac{1}{2} e^{4f}e^{-2K}W^{-2}{\mathcal{Q}}^{'2}\label{5.4}\\
T_{3}^{3}&=&\frac{1}{2}e^{-2K}{\mathcal{P}}^{'2}+\frac{1}{2}
e^{4f}e^{-2K}W^{-2}{\mathcal{Q}}^{'2}\label{5.5}
\end{eqnarray}
It is clear from the above equations that
\begin{equation}\label{5.6}
{T}^{0}_{0}+{T}^{3}_{3}= 0
\end{equation}
Drawing analogies by comparing this equation  with (\ref{1.35}) we equate
the right hand sides of equations (\ref{1.22}) and (\ref{5.2}-\ref{5.5})
to arrive at
\begin{equation}
{\mathcal{Q}}={\rm constant}\label{5.7}
\end{equation}
obviously the equation of state (\ref{1.015}) is also satisfied.
Therefore dark energy and dark pressure in terms of the vector potential are given by
\begin{equation}\label{5.8}
{\textit{U}}=-{P\over 2}=-{1\over2}\kappa^{-4}e^{-2K}{\mathcal{P}}^{'2}
\end{equation}
or converesely the solutions (26-32) are soultions of the Einstein-Maxwell equations
with the electromagnetic field given by the vector potential (33) with
\begin{equation}\label{5.9}
{\mathcal{P}}=\kappa^{2}\int e^{K}\sqrt{P(\rho)}{\rm d}\rho
\end{equation}
and ${\mathcal{Q}}$ a constant which could be taken to be zero. It is noted from 
(\ref{5.8}) that the dark energy, as expected from equations (29), (31) and (32), 
is a negative quantity. It is
shown, by calculating the tidal acceleration on the brane in the off-brane direction,
that negative dark energy enhances the localization of the gravitational field near 
the brane \cite{Maar1}.
%%%%%%%%%%%%%%%%%%%%%%%%%%%%%%%%%%%%%%%%%%%%%
\section{Discussion and summary}
we have investigated exact static cylindrically symmetric solutions
of the modified Einstein field equations for the induced metric on the brane. It is shown that, 
as in the case of the Reissner-Nordestrom type solutions \cite{Dad}, for 
the special choice of the equation of state $2U+P=0$ for the dark energy 
and dark pressure, the solutions found are that of 
the Einstein-Maxwell eqautions in the usual 4-dimensional general relativity. 
Of course there are no electromagnetic fields
present, instead, using the analogy with the solutions to the Einstein-Maxwell solutions, the 
fields present could be interpreted as 
tidal fields (\ref{5.9}) arising from the imprints, on the brane, 
of the bulk free gravitational field. 
The dark energy corresponding to the bulk free gravitational field is
shown to be negative in all the solutions obtained, indicating that it acts in favour 
of confining the gravitational field near the brane. 
Finding an exact bulk solution that reduces to the exact induced metric on the brane 
is not an easy task and simple embeddings of the 4-d solution into the 5-d equations either results in equations which could not be solved or change the nature of the 4-d solution. As an example it could be seen that the solution given by equations (28) and (29) when embedded into the Randall-Sundrum \cite{RS2} general form 
$$ds_{bulk}^2 = f(y)({dy}^2 + {ds}_{brane}^2)$$
satisfies the bulk equations {\it only} when either $c_1=0$  or $c_2=0$. But in either case the 4-d solution reduces to the Levi-Civita's {\it Ricci flat} solution \cite{Steph}. This is already known to satisfy the 
bulk equations since the Minkowski metric in Randall-Sundrum AdS solution could be replaced by {\it any} Ricci flat solution \cite{Chamb}.  
%%%%%%%%%%%%%%%%%%%%%%%%%%%%%%%%%%%%
\section*{Acknowledgement}
M. N-Z thanks University of Tehran for supporting this project under the grants provided 
by the research council.

%--------------------------------------------begin bibliography--------------------------------------
\bibliographystyle{amsplain}

\end{document}